# Dynamic ligand control of crystalline phase and habit in cesium lead halide nanoparticles

Thumu Udayabhaskararao[1], Miri Kazes,[1*] Lothar Houben,[2] Ayelet Teitelboim[1] and Dan Oron[1*]

   1 Department of Physics of Complex Systems, Weizmann Institute of Science, Rehovot 76100, Israel

   2 Chemical Research Support, Weizmann Institute of Science, Rehovot 76100, Israel

   *e-mail: miri.kazes@weizmann.ac.il ; dan.oron@weizmann.ac.il

**Active control over the shape, composition and crystalline habit of nanocrystals is a long sought-after goal. Various methods have been shown to enable post-synthesis modification of nanoparticles, including the use of the Kirkendall effect, galvanic replacement and cation or anion exchange, all taking advantage of enhanced solid-state diffusion on the nanoscale.[1,2,3,4,5] In all these processes, however, alteration of the nanoparticles requires introduction of new precursor materials. Here we show that for cesium lead halide perovskite nanoparticles, a reversible structural and compositional change can be induced at room temperature solely by modification of the ligand composition in solution. Reversible transformation of cubic $CsPbX_3$ nanocrystals to rhombohedral $Cs_4PbX_6$ nanocrystals is achieved by controlling the ratio of oleylamine to oleic acid capping molecules. This scheme does not only enable fabrication of high purity monodisperse $Cs_4PbX_6$ nanoparticles with controlled sizes. Rather, dependent on the $Cs_4PbX_6$ nanoparticles final size afforded by reaction time, the back reaction yields $CsPbX_3$ nanoplatelets with controlled thickness. These results shed some light on the unique properties of this family of materials, provide hints to the origin of their superb photovoltaic performance and point at possible routes for their controlled fabrication and stabilization.**

Hybrid organic-inorganic lead halide perovskites and their inorganic counterparts have attracted tremendous attention in the last few years, predominantly due to their superb performance as solar cell materials, but also due to potential applications in light emitting diodes and optical gain media. While studies on this family of materials (particularly the inorganics), including elucidation of their phase diagrams, have been carried on for nearly fifty years, work on colloidal nanocrystals has only recently been initiated. Here we focus on the all-inorganic cesium lead halide nanocrystals, but the tools and methods are likely applicable also to hybrid organic-inorganic materials.

The phase diagram of cesium lead halides enables to stably grow materials with several ratios of CsX and $PbX_2$.[6] A 1:1 ratio yields the most commonly studied form $CsPbX_3$ featuring a three-dimensional (3D) network of corner-shared lead halide octahedra. A 1:2 ratio yields $CsPb_2X_5$, a two-dimensional (2D) layered perovskite derivate obtained from the 3D analog by slicing along crystallographic planes and insertion of $PbX_2$ planes. Finally, a 4:1 ratio yields $Cs_4PbX_6$, a quasi zero-dimensional (0D) perovskite derivate with a recurring motif of isolated lead halide octahedra, exhibiting highly localized optical excitations and a bandgap in the near UV region.[7] As the growth conditions required to achieve either of these phases do not significantly differ, and since thin film samples are often shown to exhibit domains with different composition, it is tempting to consider the growth and transformation of cesium lead halide nanoparticles in solution. To date, reports have focused on $CsPbX_3$ nanoparticles. Yet, as we show below, these can be readily transformed into $Cs_4PbX_6$ particles with small changes to the nanoparticle chemical environment.

The route for the structural transformation between $CsPbX_3$ and $Cs_4PbX_6$ is illustrated in Fig. 1a. A small modification in the oleic acid (OA) to oleylamine (OLAm) ratio in the synthesis that produces cubic $CsPbX_3$ nanocrystals (NCs) leads to the formation of $Cs_4PbX_6$ NCs. When the ligand environment is modified from 1:1 OA to OLAm volume ratio in the initial procedure to 1:1.1, a complete transformation of the cubic phase into another phase occurs within 2-5 hours. Evidence for the conversion is the color change from green for $CsPbBr_3$ and red for $CsPbI_3$ luminescent solutions, respectively, to weakly luminescent yellow-white solutions (Figures S1-S2). The conversion time is reduced in the case of $CsPbI_3$ than $CsPbBr_3$. The gradual reduction in the absorption and fluorescence intensity of $CsPbX_3$ NCs is accompanied by the emergence of new strong absorption features in the near-UV (Figure 1b). Time dependent X-ray diffraction (XRD) spectra (Figure 1c), show that the disappearance of cubic phase $CsPbX_3$ is followed by the formation of $Cs_4PbX_6$. Despite the rather broad size distribution of the original $CsPbBr_3$ NCs, the transformed phase consists of highly monodisperse $Cs_4PbBr_6$ NCs. Transmission electron microscopy (TEM) images show that during the conversion the cubic morphology of $CsPbBr_3$ NCs changes to a rhombohedral one (Figure 1a).

Intriguingly, this conversion process to $Cs_4PbX_6$ NCs can be reversed by various methods such as the addition of oleic acid, subjection to heat (90 to 180 °C) or removal of passivating ligands. As shown in figure 1d, following the addition of oleic acid, we observed the conversion back to $CsPbX_3$, and were able to repeat this cycle simply by repeated additions of OLAm and OA up to two conversion cycles (Figure 1d). We suspect that this transformation is controlled by the OLAm to OA Brønsted acid-base type equilibrium[8,9] shifting the carboxylate to ammonium concentrations. Since multiple additions involve a decrease of the ion concentrations over time, the number of times this cycle could be continued is finite. Notably, however, washing and re-dissolution in a hexane/oleic acid mixture could produce back the $CsPbBr_3$ NCs along with remnant $Cs_4PbBr_6$ NCs. Upon addition of lost lead salts, a pure $CsPbBr_3$ phase could be reproduced (see experimental section for details). Remarkably, irrespective of the polydispersity in the starting $CsPbBr_3$ NCs the obtained $Cs_4PbBr_6$ NCs are very monodisperse.

The conversion from $CsPbBr_3$ to $Cs_4PbBr_6$ involves a dramatic change not only in crystal structure but also must be accompanied by a vast change in atomic composition. Li et al. have demonstrated growth of $CsPb_2Br_5$ sheets from orthorhombic $CsPbBr_3$ intermediates in the presence of excess $PbBr_2$ during oriented attachment.[10] This mechanism however, cannot explain the back reaction observed in the system described here. Here, we believe that at the conditions of this synthesis, modification of the solution environment leads to a change in the thermodynamically favored phase of the cesium lead halide. Surprisingly, in contrast with most colloidal systems, the kinetics of the conversion process is fast, on the order of seconds to minutes or hours even at room temperature.

We consider two possible transformation mechanisms: The first is solid phase diffusion induced either by metal-ligand complex formation or by interfacial dynamics of capping ligands and the effect of surface energy minimization.[11] However, this mechanism is unlikely because it requires a crystallographic transformation from a cubic to rhombohedral space group, including a transformation of both cation and anion sublattices. Such a transformation should require structural similarity of the both crystallographic structures into the bulk. Yet, no common interface arrangement between the sublattices of the cubic and the rhombohedral structure exists.

For cubic $CsPbBr_3$ the Br atoms are shared by two adjacent $[PbBr_6]^{1-}$ octahedra that form a 3D array where the negative charges are compensated for by the $Cs^+$ cations occupying the 12-fold coordinated sites between the octahedra.[12,13,14] $Cs_4PbBr_6$ consists of isolated $[PbBr_6]^{4-}$ octahedra forming a 0D perovskite structure, made up of ($CsPbBr_6$) chains along the c-direction separated Cs atoms. Within the chains, the $[PbBr_6]^{4-}$ octahedra are linked by common faces to triangular prisms, which are occupied by Cs atoms.[15,16] Tilting of octahedra can change the M−X−M angle from an ideal 180° to as low as ~150°, below which the structure type changes or becomes amorphous.[17,18] The disentanglement of octahedra and the separation with additional Br in all three dimensions means that a complete breakdown of the structure is necessary in order to go from one structure to the other. Furthermore, the absence of the mixed crystallographic structure in HRTEM images contradicts the notion of direct nucleation and growth of one phase out of the other.

The second mechanism is exfoliation[19,20] driven by amine intercalation[21,22], followed by recrystallization induced by micelle formation[23] or soft ligand templating[24,25,26]. This mechanism is supported by the observation of intermediate stages in TEM images showing both cubic $CsPbX_3$ NCs and $Cs_4PbX_6$ NCs along with lamellar, thin sheets and platelets and also amorphous material coexisting in solution (Figure S3). Moreover, monitoring the evolution of the emission spectrum over the course of the transformation reaction from, for example, $Cs_4PbBr_6$ to $CsPbBr_3$, different peaks emerge and disappear at discrete wavelengths corresponding to 1-5 perovskite MLs (SI movie(1)). This suggests an exfoliation process aided by the excess of amines followed by the ionic sphere rearrangement and recrystallization.[27] Further support for this mechanism is provided through the back-and-forth transformation from a mixed halide $CsPbBr_{1.5}I_{1.5}$, which eventually leads to segregation into bromine-rich and iodine rich $CsPbX_3$ nanocrystals (Figure S4).

In some sense the perovskite nanoparticles can be viewed in the more general context of phase change materials[22] where both crystalline and amorphous phases exist and recrystallization is effected by the formation of disordered clusters of particularly favorable surface tension and / or surface energy.[27]

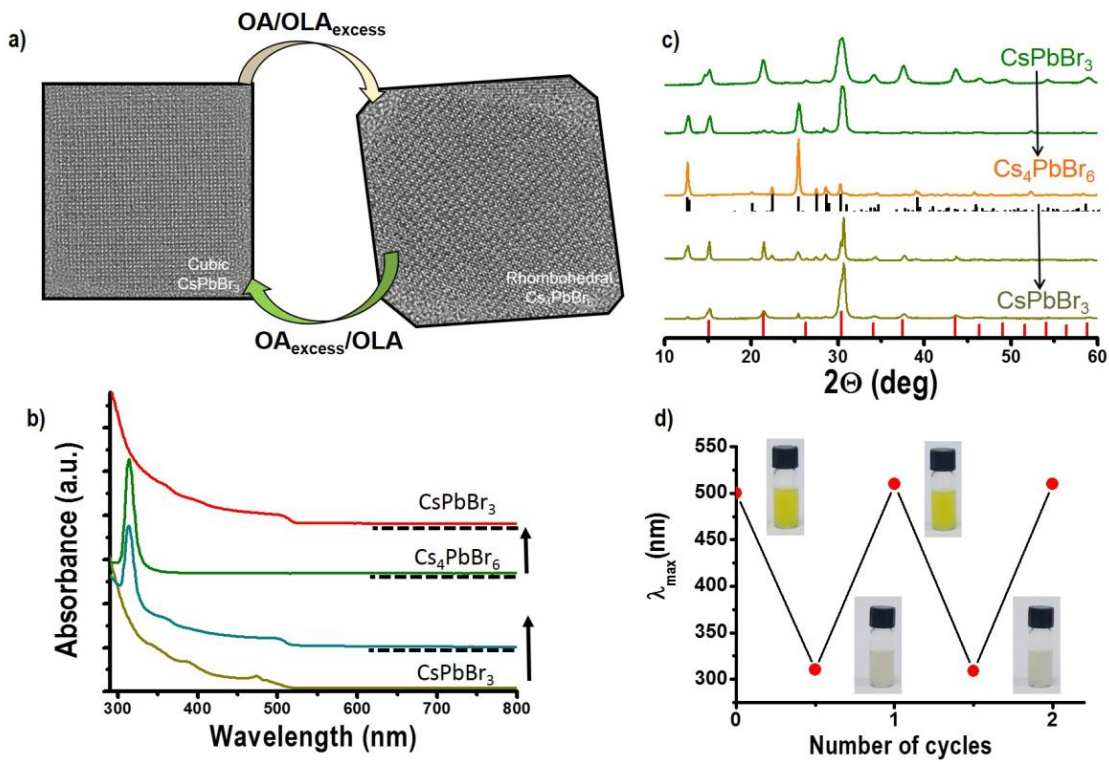

**Figure 1 | Ligand control of dynamic reversibility between $CsPbBr_3$ and $Cs_4PbBr_6$. a**, Transmission electron microscopy (TEM) images of cubic $CsPbBr_3$ to $Cs_4PbBr_6$ NCs (*left* to *right*) by means of addition of oleic acid/oleylamine excess for the forward reaction and the addition of oleic acid for the backward reaction. **b**, X-ray diffraction (XRD) of the conversion of $CsPbBr_3$ to $Cs_4PbBr_6$ NCs during the progress of reaction. **c**, changes in the absorption spectra during the progress of the forward reaction and the conversion to $CsPbBr_3$ (red trace). **d,** two complete $CsPbBr_3$–$Cs_4PbBr_6$ NCs cycles as followed by UV-vis spectroscopy. Plotted is the wavelength of the first excitonic absorption peak.

To further elucidate the mechanism underlying the reversible conversion between the two nanoparticle phases, we further modified the synthesis ligand environment. When replacing the synthesis ligand mixture with pure OLAm, quasi-spherical $Cs_4PbBr_6$ NCs are obtained along with micron sized $CsPbBr_3$ NWs (Figures S5-S6). This implies that OLAm preferentially stabilizes $Cs_4PbBr_6$ NCs. Yet, OLAm alone does not support growth in a rhombohedral geometry. A similar change to quasi-spherical shape while maintaining the phase is observed when washing rhombohedral (diamond) shaped particles (Figure S7). Surface-bound ammonium ions are more labile and will detach more easily through the washing process.[9] This means that a cooperative effect of the ligand mixture plays a crucial role in determining the nanoparticle shape following the conversion.

Size control can be achieved by a direct synthesis method where the final particle size is determined by the amount of OA/OALm used and temperature (Figure 2). Following CsOA injection at 150 °C, an inhomogeneous mixture of $Cs_4PbBr_6$ NCs is obtained. A sequential ageing step of 5 hours will result in a homogenous sample of 13 nm rhombohedral $Cs_4PbBr_6$ NCs. Ageing at increased temperature in the range of 50 – 100 °C will result in NCs of increased sizes of 23, 40 and 70 nm size particles (Figure 2a-d). Similar, well-defined sizes are also observed for $Cs_4PbI_6$ NCs (Figure 2e) and mixed halide nanoparticles (Fig. 2f).

The $Cs_4PbX_6$ NCs exhibit a strong absorption peak at 310 nm and 370 nm for the bromide and iodide, respectively.[28] Mixed $Cs_4PbBr_3I_3$ NCs were also synthesized by using 50:50 mole ratio of the corresponding lead salt in the reaction mixture, exhibiting an excitonic absorption at 340 nm (Figure 2g). These absorption spectra are indeed in agreement with those of bulk materials.[29] In contrast with some previous reports on thin films[15,30,31,32], these samples show weak emission in the near ultraviolet and blue spectral regions and no emission in the red or green. This points at the likely presence of sparse small domains of $CsPbX_3$ in $Cs_4PbX_6$ films which act as deep traps for excited carriers and dominate the emission spectrum, and is in agreement with the above mentioned notion of a thermodynamically controlled conversion process (whereby in bulk films a minority of domains appears due to the low energy difference between the two phases). The unique optical properties of $Cs_4PbX_6$ NCs make them potentially attractive for various applications, particularly ultraviolet luminescence and solar blind detection.

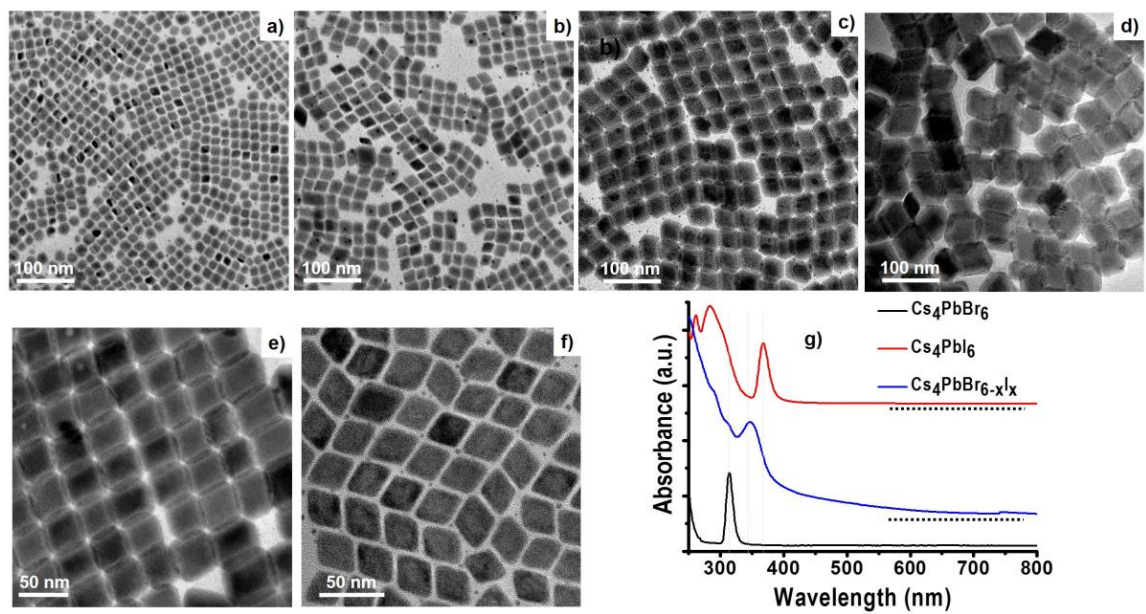

**Figure 2 | Size controlled synthesis of $Cs_4PbX_6$ NCs. a-d** are TEM images of $Cs_4PbBr_6$ NCs of sizes 14±0.7, 23±0.8, 40±1 and 70±3 nm. **e** TEM image of 50nm $Cs_4PbI_6$ NCs **f** TEM image of mixed halide 50nm $Cs_4PbBr_{1.5}I_{1.5}$ NCs. **g** absorption spectra of $Cs_4PbBr_6$, $Cs_4PbBr_3I_3$ and $Cs_4PbI_6$ NCs showing a clear shift of the excitonic peak with composition. Further data can be found in Figures S8-S10.

Reversible phase switching from $Cs_4PbX_6$ to $CsPbX_3$ at room temperature enables a new degree of control over the synthesis of $CsPbX_3$. The obtained $CsPbX_3$ NCs are highly stable as compared with typical synthesis procedures. Indeed, cubic $CsPbI_3$ NCs obtained through this procedure were stable for several months. Moreover, we could fabricate highly quantized $CsPbBr_3$ NPLs with thickness various from 1 to 5 monolayers upon the addition of OA to the dependent on $Cs_4PbBr_6$ nanocrystals size evident by the absorption and PL spectra shown in figure 3 which is in full agreement with previously reported emission peaks.[33]

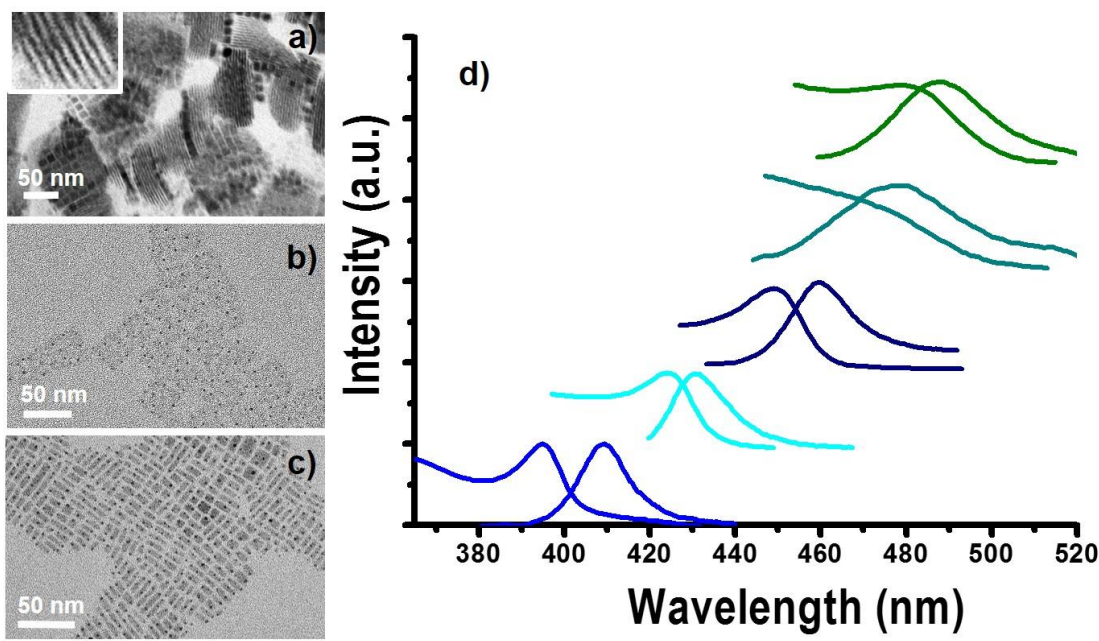

**Figure 3 | Selective transformation of $Cs_4PbBr_6$ NCs to $CsPbBr_3$ NPLs. a-c** TEM images of the $CsPbBr_3$ samples emitting at 395nm, 432nm and 490nmm respectively. **d** Absorption and emission spectra of $CsPbBr_3$ NPLs. Five different absorption and emission peaks correspond to five different thicknesses of the plates (1−5 unit cells)

In conclusion, we show here the impact of delicate changes in ligand environment on the stoichiometry and crystal structure of perovskites. This is in line with our previous work where we showed that small changes in the oleate to ammonium ratio generated by the addition of a Lewis base promotes the transition from cubic to orthorhombic $CsPbI_3$.[34] Here we show how a robust reversible transformation from $CsPbX_3$ to the rhombohedral $Cs_4PbX_6$ can be obtained via control of the OA to OLAm Brønsted acid-base type equilibrium. The extremely rapid kinetics of these transformations strongly support the notion that the lattice of lead halide perovskites is highly labile, which has been linked to the superior photovoltaic performance. Another proposition which we raise here is that due to the thermodynamic equilibrium, defects in CsPbI$_3$ films may nucleate nanodomains of Cs$_4$PbI$_6$ (or, alternatively, CsPb$_2$I$_5$) which, due to their high band gap cannot act as effective recombination centers. All in all, our results highlight new possibilities for engineering lead halide based perovskites based via control of their surface properties.

**Author contributions**

U.B. synthesized the nanocrystals and performed XRD and TEM measurements. L.H. performed HRTEM measurements. U.B., A.T. and M.K. performed the optical characterization. M. K. and D.O. conceived and supervised the project. The manuscript was jointly written by all authors.

**Additional information**

Correspondence and requests for materials should be addressed to D.O.

**Acknowledgements**

This research was supported by a grant from Israel Science foundation (grant no. 2012\224330*) and by the Crown Center of Photonics and the ICORE: the Israeli Excellence Center "Circle of Light". The support of the Gerhardt M.J. Schmidt Minerva Center and the Irving and Cerna Moskowitz Center for Nano and Bionano Imaging and at the Ernst-Ruska Centre Juelich, Germany, is also gratefully acknowledged.

**Methods**

**Nanoparticle synthesis:**

**Chemicals:** All chemical reagents were purchased from commercial suppliers and used without purification. $Cs_2CO_3$ (99.9%), octadecene (ODE, 90%) and oleic acid (OA, 90%), $PbI_2$ (99.999%), $PbBr_2$ (98%) were purchased from Sigma-Aldrich. Oleylamine (OLAm, 80-90%) was purchased from Acros.

**Preparation of Cs-oleate:** $Cs_2CO_3$ (0.814g) was loaded into 100 mL 3-neck flask along with ODE (40mL) and OA (2.5 mL), dried for 1h at 120 ºC, and then heated under argon to 150 ºC until all $Cs_2CO_3$ reacted with OA.

**Synthesis of cubic $CsPbX_3$ NCs:** ODE (5 mL) and $PbX_2$ (0.188 mmol) such as $PbI_2$ (0.087g) and $PbBr_2$ (0.069g) were loaded into 25 mL 3-neck flask and dried under vacuum for 1h at 120 ºC. Dried OLAm (0.5 mL) and dried OA (0.5 mL) were injected at 120 ºC under argon flow. After various ageing times the temperature was raised to 170 ºC and CsOA solution (0.05 mmol, 0.4 mL, 0.125 M in ODE, pre-heated to 100 ºC) was quickly injected and 5s later, the reaction mixture was cooled by the ice-water bath.

**Purification of the $CsPbX_3$ NCs:** The crude reaction mixture was centrifuged at 6000 rpm for 5 min. The supernatant was separated and 20 mL ethyl acetate was added as an anti-solvent (the volume ratio of the original supernatant to anti-solvent is about 1:4). The clear supernatant solution immediately turns cloudy centrifuged. The precipitate is used for the next stage.

**Transformation from $CsPbX_3$ NCs to $Cs_4PbBr_6$ NCs:** OA and OLAm (;The addition of 0.4 ml of 1ml OLAm and 0.9ml OA dissolved in 1 ml hexane were added to half of the crude perovskite batch resulted in a slow transformation to $Cs_4PbBr_6$ over a 2 day period. The addion of 0.8ml accelerated the process to ~10h. The sample was stirred until the color changes from their respective color to yellow white solution, indicating the formation of $Cs_4PbX_6$. Sequential addition of 0.4 mL (2mmol) of OA to the above solution results in the appearance of $CsPbX_3$. This cycle can be repeated for two more time. However, washing and re-dissolution of the $Cs_4PbBr_6$ NCs in a hexane/oleic acid mixture could produce back the CsPbBr$_3$ NCs along with remnant Cs$_4$PbBr$_6$ NCs. Upon addition of 50uL PbX$_2$ (0.188 mmol) dissolved in 5ml ODE, a pure CsPbBr$_3$ phase could be reproduced.

**Cs$_4$PbBr$_6$ NCs Size control:** Cs$_4$PbBr$_6$ NCs size control is achieved by modification of for CsPbBr$_3$ described above.

ODE (5 mL) and PbBr$_2$ (0.069g) were loaded into 25 mL 3-neck flask and dried under vacuum for 1h at 120 °C. Dried OLAm (0.8 mL, 4.36 mmol) and dried OA (0.7 mL, 3.6mmol) were injected at 120 °C under argon flow. Then the temperature was raised to 150 °C and CsOA solution (0.05 mmol, 0.4 mL, 0.125 M in ODE, pre-heated to 100 °C) was quickly injected. The reaction mixture was cooled 5s later by immersion in an ice-water bath. This reaction was continued at room temperature for 2-5 hours resulted in the formation of 13nm Cs$_4$PbX$_6$ NCs as indicated by the color changes from green to pale yellow. The crude reaction mixture was centrifuged at 6000 rpm for 5 min without anti-solvent.

25 nm Cs$_4$PbBr$_6$ NCs was produced by using 0.7 mL OLAm (3.82mmol) and 0.6ml OA (3.1 mmol) and continuing the reaction at 40 °C for 5-10 hours. Continuing the reaction at 60 °C results in the formation of 50 nm and 70 nm Cs$_4$PbBr$_3$ NCs at different time intervals.

**Thickness control of CsPbBr$_3$ NPLs through the reverse reaction from Cs$_4$PbX$_6$ NCs:** Synthesis of CsPbBr$_3$ NPLs whose emission are 408, were made form 13 nm quasi-spherical Cs$_4$PbBr$_6$ NPs upon dilution of these NCs with 0.1 mmol OA contained in hexane. NPLs with 432 nm emission were obtained from crude (unwashed) 13 nm rhombohedral NCs. Further addition of 20 uL oleic acid shifts the emission to 430 nm. NPLs of 460 and 480 nm emission peaks were produced from 25 nm Cs$_4$PbBr$_6$ NCs And an addition of 100-200 uL OA

**Optical measurements:** UV-vis absorption spectra were measured using a UV-vis-NIR spectrometer (V- 670, JASCO). Fluorescence spectrum was measured using USB4000 Ocean Optics spectrometer excited by a fiber coupled 407 nm LED in an orthogonal collection setup.

**Transmission Electron Microscopy (TEM) imaging:** TEM studies were carried out on a CM120 Super Twin microscope (Philips) operating at 120 kV. High-resolution images were taken at an acceleration voltage of 80 kV on the spherical and chromatic aberration-corrected FEI 60-300 Ultimate ('PICO') instrument at the Ernst Ruska-Centre, Germany.[35]